\documentclass{ws-ijmpb}

\begin{document}

\markboth{B. K. Raj, B. Pradhan and G. C. Rout}
{Modified BCS gap equation for Jahn-Teller distorted......}

%
\catchline{}{}{}{}{}
%

\title{MODIFIED BCS GAP EQUATION FOR JAHN-TELLER DISTORTED
 HIGH TEMPERATURE SUPERCONDUCTORS }

\author{B. K. RAJ}

\address{Department of Physics, Government (Autonomous) College,\\
 Angul-759128, India}

\author{B. PRADHAN\footnote{Corresponding author, Email: brunda@iopb.res.in, Mob: +91-9437806565}}

\address{Department of Physics, Government Science College,\\
 Malkangiri-764048, India}

\author{G. C. ROUT}

\address{Condensed Matter Physics Group,
P. G. Department of  Physics,\\
 F. M. University, Balasore-756019, India}
\maketitle

\begin{history}
\received{xxx}
\revised{xxx}
\end{history}

\begin{abstract}
In this communication we report the interplay of the normal 
electron-phonon (EP) interaction, dynamic Jahn-Teller (DJT) distortion 
and superconductivity in high temperature superconductors in presence 
of a static lattice strain. This model consists of a degenerate two 
orbital band separated by Jahn-Teller (JT) energy  modified by the DJT 
interaction in  the conduction band. The superconductivity is assumed to
 be $s$-wave type present in the same band. The interaction Hamiltonian is 
solved by Green's function method and a modified BCS gap equation is 
obtained with a modified conduction band energy $\tilde{\epsilon}_{\alpha k}$ and
 modified BCS order parameter $\tilde{\Delta}_{\alpha}$ with $\alpha=1,~2$
 designating the two orbitals. This interplay displays some new interesting 
results which are different from the effect of the static lattice strain
 on superconducting (SC) order parameter. The interplay is studied by varying the normal
 EP coupling, the DJT coupling, the SC coupling, the phonon vibration 
frequency, the phonon wave vector and other model parameters of the system.
\end{abstract}


\section{Introduction}
The high-T$_c$ copper oxide superconductors exhibit a pseudogap (PG)
 state\cite{int1,int2,int3} as observed in the anomalous transport\cite{int4}, 
thermodynamic\cite{int5}, and optical\cite{int6} properties below
 a temperature, larger than the superconducting transition
temperature T$_c$. One of the challenging issues of the SC mechanism
 is the origin of the PG. The PG phase is detected in the underdoped region,
 which has been interpreted in terms of preformed pairs\cite{int7},
 a circulating current phase\cite{int8,int9,int10} and many more.
In many cuprate superconductors\cite{int11,int12} in the underdoped 
region, the static stripe phases play an important role, which may be
 incommensurate unidirectional spin and charge-density 
waves\cite{int13}. Also it has been reported that 
bismuth-based cuprates conventionally 
 exhibit the "out-of plane disorder," which 
 strongly affects superconducting 
transition temperature T$_c$\cite{int15,int16}. This suggests that the out-of
 plane disorder is a potentially important factor concerning the 
superconducting properties of high-T$_c$ superconductors 
and hence it proves the importance of Jahn-Teller effect.

The dependence of the SC transition temperature on the
 structural properties is little understood for the high temperature SC
 compounds. In this system, the Fermi level (FL) lies within the degenerate 
band of the two orbitals. The molecular distortion is produced when the 
electronic degeneracy is removed by some symmetry breaking interactions.
 This type of distortion is called Jahn-Teller distortion. There is a
 change in the electronic density of states (DOS) around the FL associated
 with such a transition. Therefore, it is expected that the band Jahn-Teller
 (BJT) distortion would strongly influence the superconductivity in such a
 system.

There is a large number of experimental evidences indicating the 
strong influence of the structural distortion on the SC transition 
temperature. The neutron scattering measurements of the temperature
 dependence of the spontaneous strain have been reported by Paul
 $et~al.$\cite{int16a} for $La_{2-x}Ba_xCuO_4$ ($T_c\simeq 38 K$). 
The system exhibits
 a structural transition at $180K$. The magnitude of the strain rises on
 lowering the temperature and shows an anomalous suppression below $75K$. This 
anomaly is expected to be associated with the appearance of the 
superconductivity in the system. The evidence for
 such interplay between lattice 
distortion and superconductivity in $La_{2-x}Sr_xCuO_4$, $YBa_2Cu_3O_{6.5}$ 
and even in electron doped cuprate system $Nd_{2-x}Ce_xCuO_4$ has been 
reported by Lang $et~al.$\cite{int16b} from thermal expansion measurements.
 Moreover, the suppression of the lattice distortion in the SC state has also
 been observed for $2212$ and $2223$ $Bi$-superconductors as shown by 
extended X-ray absorption measurements\cite{int16c}. Further, the structural
 transition from tetragonal to orthorhombic phase in $La_{2-x}Sr_xCuO_4$ takes
 place at temperatures higher than the SC transition temperature. The doping
 dependence of the transition temperature T$_c$ is also reported for 
$La_{2-x}M_xCuO_4$ (M~=~Ca, Sr)\cite{int16d}. These measurements clearly
 demonstrate that there exists a strong interplay between superconductivity 
and structural distortion.

The introduction of an electron-phonon mechanism that can 
achieve high-T$_c$ superconductivity is challenging in the 
theoretical point of view. For the theoretical study of the 
high-T$_c$ superconductivity the onsite and/or intersite
bipolarons, JT bipolarons and different mechanisms\cite{int17,int18} of
carrier dynamics such as Bose-Einstein condensation and
tunneling-percolation have been proposed. The JT  polaron pairing
 effect was originally
proposed as a possible explanation for the superconductivity
in $La_{2-x}Ba_xCuO_4$ by Bednorz and Mu$\ddot{u}$ller\cite{int19}. 
Since then this effect has been discussed by a number of authors in 
different contexts\cite{int20,int21,int22}  and many features have 
been observed experimentally supporting the general concept of JT 
effect\cite{int23,int24}. A dynamic Jahn-Teller
  effect arises, when there is a degeneracy in both of electronic
levels and of molecular vibrations and
when two independent distortions have the same JT energy
lowering. The DJT effects have
been proposed to play a role in a number of materials of
current interest including cuprates\cite{int21,int25,int27} and
manganites\cite{int28} etc.

Ghosh $et~al.$ have reported a simple model study of the interplay 
between the superconductivity and the static JT distortion\cite{int29}. 
In another report the same authors have reported the coexistence of the 
static JT effect with the superconductivity for the correlated orbitally 
degenerated bands of the cuprate systems taking into account one band 
Hubbard model\cite{int30}. In the present report, we consider the interplay 
between static lattice strain and superconductivity in presence of a normal
 electron-phonon interaction and dynamic JT interaction.
The rest of the work in the paper is organized as follows. The 
theoretical model for the DJT is described in section 2. In the dynamic
 limit calculation, the electron Green's functions are calculated and
 solved in such a way that it gives rise to the modified BCS gap equation.
 The lattice strain relation 
is calculated in section 3, the results and discussion are presented in
 section 4 and finally the conclusion is given in section 5.

\section{Theoretical Methods}
The present model study attempts to investigate the effect of DJT
 distortion on the superconducting  gap in high-T$_c$ 
superconductors (HTSs). If the Fermi level lies on a degenerate conduction
 band and the degeneracy is removed by lowering of the lattice symmetry, 
a spontaneous lattice strain is produced in the band resulting in the 
structural transition. This strain in the band is stabilized, if the gain 
in electronic energy overcomes the cost in elastic energy. Depending on 
the magnitude of the lattice distortion, the system can behave as a metal 
or insulator. The difference in the occupation probability of the two bands
 will couple to the lattice strain which gives rise to JT
 distortion. The Hamiltonian for such a system is reported
 earlier\cite{mod1,mod2,mod3} and the same Hamiltonian is written here.
\begin{equation}
H_c=\sum_{k\sigma}\epsilon_{k}\left(c^{\dagger}_{1k\sigma}
 c_{1k\sigma}+c^{\dagger}_{2k\sigma}c_{2k\sigma}\right),
\label{ee1}
\end{equation}
\begin{equation}
H_{e-L}=Ge~\sum_{k\sigma}\left(c^\dagger_{1k\sigma}c_{1k\sigma}
- c^\dagger_{2k\sigma}c_{2k\sigma}\right).
\label{ee2}
\end{equation}
The Hamiltonian $H_c$ represents the hopping of the electrons between 
the two nearest neighbors for the two degenerate orbitals designated 
as 1 and 2. The dispersion of the degenerate band in a two dimensional
 CuO$_2$ plane is written as $\epsilon_k=-2t_0(\cos k_x+\cos k_y)$.
 Here $c^{\dagger}_{\alpha k\sigma}(c_{\alpha k\sigma})$, for $\alpha=1$ 
and 2, are creation (annihilation) operators of the conduction electrons
 of copper ions for two orbitals with momentum $k$ and spin $\sigma$. 
The Hamiltonian $H_{e-L}$ represents the static JT interaction where $G$
 is the strength of the electron-lattice interaction and $e$ is the strength 
of the isotropic static lattice strain. The lattice strain splits the 
single degenerate band into two bands with energies 
$\epsilon_{1,2k}=\epsilon_{k}\pm{Ge}$. The elastic energy of the system is
 $\frac{1}{2}Ce^{2}$ with $C$ representing the elastic constant. The 
minimization of the free energy of the electron including the
 elastic energy helps to find the expression for 
lattice strain. It is shown earlier\cite{mod1,mod2,mod3} that the static 
lattice strain suppresses the SC gap parameter in the interplay region of 
the two order parameters.

In order to investigate the phonon response in the HTSs,
  we consider the phonon interaction to the density
 of the conduction electrons of both the orbitals as well as the phonon 
coupling to the difference in electron densities of the JT distorted 
 orbitals of the conduction band. The electron-phonon interaction 
Hamiltonian is written as 
\begin{eqnarray}
H_{e-p}&=&\sum_{\alpha,k,\sigma}f_1(q)\left(c^{\dagger}_{\alpha k+q \sigma}
c_{\alpha k \sigma}\right)A_{q}-\sum_{\alpha,k,\sigma}(-1)^{\alpha}f_2(q)e\left(c^{\dagger}_{\alpha k+q \sigma}
c_{\alpha k \sigma}\right)A_{q}\nonumber \\
&=&\sum_{\alpha, k,\sigma}s_{\alpha}(q)\left(c^{\dagger}_{\alpha k+q \sigma}
c_{\alpha k \sigma}\right)A_{q}.
\label{ee3}
\end{eqnarray}
The strength of the EP coupling $s_{\alpha}(q)$ is defined as
 $s_{\alpha}=f_1(q)-(-1)^{\alpha}f_2(q)e$ in which $f_1(q)$ is 
the normal EP coupling and $f_2(q)$ is the dynamic Jahn-Teller EP coupling. 
The $q^{th}$-Fourier component of the phonon displacement operator is
 $A_{q}=b_{q}+b_{-q}^{\dagger}$ with $b_{q}^{\dagger}$ ($b_{q}$) 
defining the phonon  creation (annihilation) operator for wave vector $q$. 
Further the free phonon Hamiltonian $H_p$ is given in harmonic approximation as
\begin{equation}
H_p=\sum_{q}\omega_{q}b^{\dagger}_{q}b_{q},
\label{ee4}
\end{equation}
with $\omega_{q}$ being the free phonon frequency.

Our main objective in the present report is to study the effect of
 dynamic JT distortion on the superconductivity in HTSs.
 The $d$-wave models have gained substantial support recently
over $s$-wave pairing as the mechanism by which high temperature
superconductivity might be explained. The establishment of $d$-wave
symmetry in cuprates does not necessarily specify a high-T$_c$
 mechanism. It does not impose well defined constraints on possible
models for this mechanism. While the spin fluctuation pairing
mechanism leads naturally to an ordered parameter with $d$-wave
symmetry, the conventional BCS electron phonon pairing interaction
give rise to $s$-wave superconductivity. As a first step, it is assumed
that $s$-wave like BCS pairing interaction mediated by some boson
exchange exist within the same orbitals of a sub lattice and the
same strength of the interaction is taken for the orbitals. 
The BCS type pairing 
Hamiltonian is considered here for the two orbitals.
\begin{equation}
H_I=-\Delta\sum_{\alpha, k}\left(c_{\alpha k\uparrow}^{\dagger}c_{\alpha, -k\downarrow}^{\dagger}+c_{\alpha, -k\downarrow}c_{\alpha k\uparrow}\right).
\label{ee5}
\end{equation}
In order to simulate an attractive interaction to produce Cooper pairs,
 the energy dependence of the interaction potential is taken as
\begin{equation}
U(\epsilon)=U_0\left[1-\frac{(\epsilon-\epsilon_F)^4}{\omega_{D}^{4}} \right]^{\frac{1}{2}},
\label{ee6}
\end{equation}
where $U_0$ is the effective attractive Coulomb interaction, $\omega_D$ 
is cut-off energy and $\epsilon_F$ represents the Fermi energy. 
In order to produce band splitting due to JT distortion, we consider an 
energy dependent density of state $N(\epsilon)$ around the center of the
 conduction band in the system. Such a logarithmic model density
 of state\cite{mod3} is given by
\begin{equation}
N(\epsilon)=N(0)\sqrt{1-|{\frac{\epsilon}{D}}|}ln|{\frac{D^2}{\epsilon^2}}|,
\label{ee7}
\end{equation}
where $2D=W$ is the conduction band width. The total Hamiltonian 
describing the dynamic JT effect and SC interaction in high-T$_c$ 
cuprate systems can be written as
\begin{equation}
H=H_c+H_{e-L}+H_I+H_{e-p}+H_p.
\label{ee8}
\end{equation}
\section{Calculation of Order Parameters}
In order to calculate the SC gap and the lattice strain, we calculate
 the Green's functions for the electrons in the two orbitals 
($\alpha=1$ and 2) of the conduction band by using the Zubarev's technique 
of Green's functions\cite{mod4}. For the calculation we use the total 
interaction Hamiltonian given in equation (\ref{ee8}). The coupled 
Green's functions for the two orbitals can be defined as
\begin{equation}
A_{\alpha}(k,\omega)=\left<\left<c_{\alpha k\uparrow}; c^{\dagger}_{\alpha k\uparrow}\right>\right>_{\omega}~~;~~B_{\alpha}(k,\omega)=\left<\left<c^{\dagger}_{\alpha,-k\downarrow}; c^{\dagger}_{\alpha k\uparrow}\right>\right>_{\omega}.
\label{ee9}
\end{equation}
The calculation of Green's functions $A_{\alpha}$ and $B_{\alpha}$ 
couples to the higher order Green's functions $\Gamma^{1}_{\alpha}$ and
 $\Gamma^{2}_{\alpha}$ involving the electron and the phonon operators.
 They are written as
\begin{equation}
\Gamma_{\alpha}^{1}(k,q,\omega)=\left<\left<c_{\alpha k-q\uparrow}A_{q};c^{\dagger}_{\alpha k\uparrow}\right>\right>_{\omega}~;~\Gamma_{\alpha}^{2}(k,q,\omega)=\left<\left<c_{\alpha k-q\downarrow}A_{q};c^{\dagger}_{\alpha k\uparrow}\right>\right>_{\omega}.
\label{ee10}
\end{equation}
The equations (\ref{ee9}) and (\ref{ee10}) involve
 higher order Green's functions
   which are truncated by applying mean-field approximation in order to
 obtain the close form solutions. By this calculation we obtain the second
 order terms in electron-phonon coupling in the final results. In this
 mean-field approximation there appears a term $N_{q}=2\nu_{q}$, where 
the Bose-Einstein distribution
      function is $\nu_{q}=(exp({w/k_{B}T)-1)^{-1}}$.
These coupled equations for $\Gamma_{\alpha}^{1}$ and $\Gamma_{\alpha}^{2}$ 
are solved in terms of the Green's functions $A_{\alpha}$ and $B_{\alpha}$ 
and they are written below as
\begin{equation}
\Gamma_{\alpha}^{1}(k,q,\omega)=s_{\alpha}N_{q}\left[\frac{(\omega-\epsilon_{\alpha k-q})A_{\alpha}(k,\omega)-\Delta B_{\alpha}(k,\omega)}{\omega^{2}+E^{2}_{\alpha k-q}-2\omega\epsilon_{\alpha k-q}}\right]
\label{ee11}
\end{equation}
and
\begin{equation}
\Gamma_{\alpha}^{2}(k,q,\omega)=s_{\alpha}N_{q}\left[\frac{(\omega-\epsilon_{\alpha k-q})B_{\alpha}(k,\omega)+\Delta A_{\alpha}(k,\omega)}{\omega^{2}+E^{2}_{\alpha k-q}-2\omega\epsilon_{\alpha k-q}}\right].
\label{ee12}
\end{equation}
On substitution of $\Gamma_{\alpha}^{1}$ and $\Gamma_{\alpha}^{2}$ in equation (\ref{ee9}), finally we obtain the Green's functions $A_{\alpha}$ and $B_{\alpha}$ as
\begin{equation}
A_{\alpha}(k,\omega)=\frac{1}{2\pi}\left[\frac{\omega+\tilde{\epsilon}_{\alpha k}}{\omega^{2}-\tilde{E}^{2}_{\alpha k}}\right]
\label{ee13}
\end{equation}
and
\begin{equation}
 B_{\alpha}(k,\omega)=\frac{1}{2\pi}\left[\frac{-\tilde{\Delta}_{\alpha}}{\omega^{2}-\tilde{E}^{2}_{\alpha k}}\right].
\label{ee14}
\end{equation}
The modified SC quasi-particle energy band for the two orbitals can be
 written as $\tilde{E}^{2}_{\alpha k}=\left(\tilde{\epsilon}^{2}_{\alpha k}+\tilde{\Delta}^{2}_{\alpha}\right)$. Again the modified conduction
 band $\tilde{\epsilon}_{\alpha k}$ can be written in general form as
\begin{equation}
\tilde{\epsilon}_{\alpha k}=\epsilon_{\alpha k}+s^{2}_{\alpha}\sum_{q}\frac{2(\omega-\epsilon_{\alpha k-q})N_q}{(\omega-\epsilon_{\alpha k-q})^{2}+\Delta^{2}},
\label{ee15}
\end{equation}
where $\epsilon_{\alpha k-q}=\epsilon_{k-q}-\mu-(-1)^{\alpha}Ge$ and 
the Bose-Einstein distribution function N$_q$ is defined as 
$N_q=(e^{\beta\omega_q}-1)^{-1},$ with $\beta=1/k_BT$ 
and $\omega_q$ being the renormalized phonon frequency at temperature T.

The modified SC gap for the orbitals 1 and 2 in presence of
normal phonon interaction as well as the dynamic JT interaction appears as
\begin{equation}
\tilde{\Delta}_{\alpha k}=\Delta+{s^{2}_{\alpha}}\sum_{q}\frac{2\Delta N_q}{(\omega-\epsilon_{\alpha k-q})^{2}+\Delta^{2}}.
\label{ee16}
\end{equation}
From equations (\ref{ee15}) and (\ref{ee16}), it appears that the degenerate 
conduction  band $\epsilon_{k}$ and the BCS gap parameter are renormalized
 by the static Jahn-Teller effect through the EP coupling constants and
 lattice strain e. In absence of EP coupling, the quasi particle  band 
can be written as $ E_{\alpha k}=\sqrt{E^{2}_{\alpha k}+\Delta^{2}},$
  where $\epsilon_{\alpha k}=\epsilon_{k}-\mu-(-1)^\alpha Ge$.
       This agrees with the calculation of Ghose $et~al.$\cite{int29}
 for static lattice strain.
          Thus, the static Jahn-Teller effect only renormalizes the energy
 of the doubly degenerate conduction electron band, but not the 
superconducting gap parameter. Due to the interplay of both the order 
parameters i.e., superconducting gap and lattice strain, the lattice 
strain changes when the temperature decreases down to below transition 
temperature $T_{c} $\cite{int29}. In the absence of lattice strain, the
 quasi particle of BCS model is reproduced i.e., 
$E_{k}=\sqrt{E^{2}_{ k}+\Delta^{2}}.$

The SC gap parameter is defined as
\[\Delta=-\sum_{\alpha k}\tilde{U_k}\left<c_{\alpha k\uparrow}^{\dagger}c_{\alpha,-k\downarrow}^{\dagger}\right>.\]
Finally the SC order parameter is found from the correlation functions 
calculated from the Green's functions given in equation (\ref{ee14}) and
 can be written as
\begin{equation}
1=\int_{-\omega_{D}}^{\omega_{D}}U(\epsilon)N(\epsilon)d\epsilon_{k}\left[\frac{1}{2\tilde{E}_{1k}}
\tanh\left(\frac{1}{2}\beta\tilde{E}_{1k}\right)+\frac{1}{2\tilde{E}_{2k}}\tanh\left(\frac{1}{2}\beta\tilde{E}_{2k}\right)\right],
\label{ee17}
\end{equation}
where $U(\epsilon)$ and  $N(\epsilon)$ are defined in equations (\ref{ee6}) 
and (\ref{ee7}) respectively and the SC coupling constant is written 
as $g=N(0)U_0$. The equilibrium value of the static stress $e$ is found 
by minimizing the free energy which includes electronic as well as the 
lattice energies. The lattice strain is defined as
\begin{equation}
e=-\sum_{\alpha k\sigma}(-1)^{\alpha}\left<c_{\alpha k\sigma}^{\dagger}c_{\alpha k\sigma}\right>.
\label{ee18}
\end{equation}
The correlation functions of the electrons of the two orbitals are 
calculated from the Green's functions $A_{\alpha}$ given in equation
 (\ref{ee13}) and the equilibrium lattice strain can be written as
\begin{equation}
e=\left(\frac{-G}{C_{0}}\right)\int_{-W/2}^{W/2}N(\epsilon)d\epsilon_{k}\left[\frac{\tilde{\epsilon}_{1k}}{2\tilde{E}_{1k}}\tanh\left(\frac{1}{2}\beta\tilde{E}_{1k}\right)-\frac{\tilde{\epsilon}_{2k}}{2\tilde{E}_{2k}}\tanh\left(\frac{1}{2}\beta\tilde{E}_{2k}\right)\right].
\label{ee19}
\end{equation}
In order to study the mutual influence of the dynamic lattice distortion 
and superconductivity, one has to solve the above two coupled equations
 (\ref{ee17}) and (\ref{ee19}) self-consistently.

\section{Results and Discussion}

The electron Green's functions $A_{\alpha}(k,\omega)$ and 
$B_{\alpha}(k,\omega)$ given in equations (\ref{ee13}) and (\ref{ee14})
 are calculated from the total Hamiltonian consisting of SC 
and DJT interactions. The SC gap equation is written in the form 
of the modified BCS gap equation from electron correlation functions. The SC
 gap $\Delta$ and the lattice strain $e$ are calculated and given 
respectively in equations (\ref{ee17}) and (\ref{ee19}). These 
equations are solved self-consistently and their temperature dependence is 
shown in figures 1 to 12. All the physical parameters in the present 
calculation are scaled by the
  nearest neighbor hopping integral $2t_{0}\approx{0.25eV}$ for convenient
    self-consistent solutions. The dimensionless parameters are the 
SC coupling constant, $g=N(0)V_0$, the static 
JT coupling, $g_1=G/2t_0$, the SC gap parameter, $z=\Delta/2t_0$, the static JT energy, 
$\tilde{e}=Ge/2t_0=g_1e$, and the phonon energy $e_q=v_Fq$ corresponding to
the electron velocity $v_F$ at the Fermi level.
 Further the other phonon parameters 
are the normal EP coupling $\lambda_1=f_1/2t_0$, the  
 DJT coupling, $\lambda_2=f_2/2t_0$ for the JT distorted two orbitals, the 
phonon frequency, $\omega_1=0.0008$ at a given reduced temperature, 
$t=k_{B}T/2t_0$ and reduced external frequency, $c=\omega/2t_0=0.1$. The 
temperature dependent parameters $z(t)$ and $\tilde{e}(t)$ are solved 
self-consistently for a standard set of parameters 
i.e., $g=0.031$, $g_1=0.152$, 
 $\lambda_1=0.004$ and $\lambda_2=0.01$. The conduction band width, $W=1eV$ and 
the cut-off frequency, $\omega_c=250K$ for the SC pairing are taken for all 
further calculation.

Figure 1 shows the temperature variation of $z(t)$ and $\tilde{e}(t)$ 
for the static JT effect (red curves) and the DJT effect
 (green curves). The parameters are so adjusted that the JT 
distortion temperature $(t_{s})$ becomes greater than the SC transition
 temperature $(t_{c})$. The neutron scattering measurement
 of the temperature dependence of the spontaneous lattice strain for
 $La_{1.85}Ba_{0.15}CuO_{4}$ $(T_{c}\approx{38K})$ exhibits a structural 
transition at $180K$\cite{int16a}. The magnitude of the strain rises
 on lowering the temperature and an anomalous suppression of strain below
 $T<75K$. It may be associated with appearance of superconductivity.
 In static case, the SC order parameter $z$ shows mean-field 
behavior, while the JT energy gap $\tilde{e}$
 shows depression within the interplay
 region; but $\tilde{e}$ shows mean-field behavior in the JT distorted phase
 for temperature $t>t_c$. Under DJT effects, the SC gap parameter is 
enhanced throughout the temperature range with the enhancement of the 
transition temperature $t_c$. This is contrary to the results obtained for 
the system under static limits as shown by mean-field calculations as well 
as the sophisticated Slave-boson calculations\cite{int30,int29,mod2}.
It is found
 that the reduced SC gap size is $2\Delta_0/k_BT_c\simeq 3.64$ in the static JT 
limit, while the gap size is $2\Delta_0/k_BT_c\simeq 3.59$
 in the DJT limit. The ratio is
 slightly less in the dynamic limit than its value in the static limit due to 
the enhancement of $t_c$ in the dynamic limit. The reduced gap sizes
 are comparable to the universal BCS value of $3.52$. The evolutions of the 
two gap parameters are discussed below by varying the other parameters of the
 system as shown in the figures from 2 to 12.
\vspace{0.2in}

\bigskip

{\vbox{
\begin{center}
      \epsfig{file=dyna1c.eps,width=8cm,height=6cm}
      \vspace{-1.0cm}
\end{center}
}}

\vspace{0.5cm}

{\small {\bf Fig.} \ {\bf 1} \
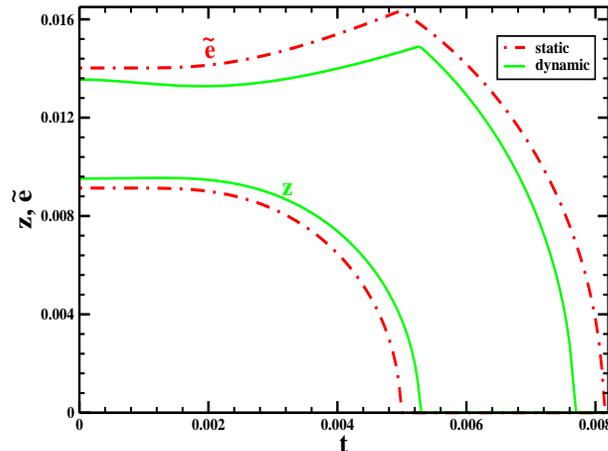
The self-consistent plots of $z$ vs. $t$ and $\tilde{e}$ vs. $t$ for the SC coupling $g=0.031$ and static JT coupling $g_1=0.152$ (red curves) and for the dynamic effect (green curves) for the same values of $g$ and $g_1$ with the fixed values of the phonon frequency $\omega_1=0.0008,$ external frequency $c=0.1,$ normal EP coupling $\lambda_1=0.004$ and DJT coupling $\lambda_2=0.01$.}

The effect of the static JT coupling $g_1$ on the SC parameter $z$ and 
the static JT energy gap $\tilde{e}$  is shown in figure 2. It is observed
 that the lattice strain energy $(\tilde{e})$ is reduced throughout
 the temperature range with 
decrease of static lattice coupling $g_1$. With the decrease of JT coupling to
 $0.145$, the mean-field behavior disappears and the gap becomes nearly
 constant with temperature. On the other hand, the SC gap is enhanced
 throughout the temperature range with 
 the decrease of the static JT coupling $g_1$ resulting in the
 enhancement of the SC transition temperature. The SC 
parameter shows mean-field behavior for all values of JT 
coupling ($g_1=0.145$  to $0.159$). It is found
 that the reduced gap size $2\Delta_0/k_BT_c$ gradually decreases with the 
decrease of the JT coupling $g_1$. This so happens because $t_c$ enhancement
 is more than the enhancement of the gap parameter $z$ due to the interplay 
of SC and lattice strain parameters. It appears that 
there exists a strong interplay between these two interactions. The band
 Jahn-Teller effect lifts the orbital degeneracy and lowers the electronic
 energy due to lattice distortion. Both structural transition $t_{s}$ and
 lattice strain $e (T=0)$ are maximum, as the Fermi level lies at the 
singular point of the density of states. The suppression of SC transition
 temperature $t_{c}$ in
  presence of structural distortion is related to the removal of electronic
 states from the Fermi level due to the lifting of degeneracy. The larger 
the splitting of orbitals (with higher values of $t_{s}$ and e), the larger
 is the suppression of $t_{c}$.
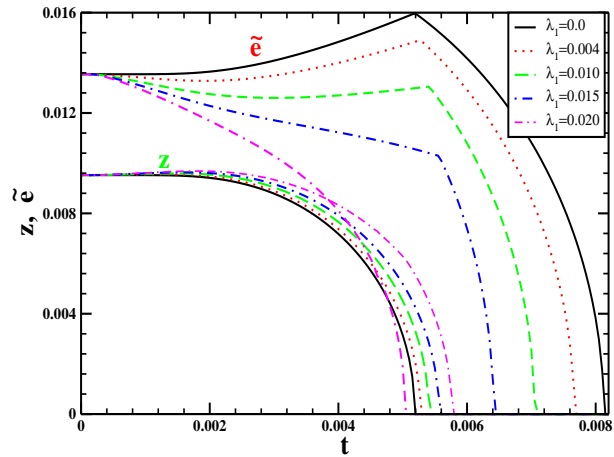
\vspace{0.2in}

\bigskip

{\vbox{
\begin{center}
      \epsfig{file=dyna3c.eps,width=8cm,height=6cm}
      \vspace{-1.0cm}
\end{center}
}}

\vspace{0.5cm}

{\small {\bf Fig.} \ {\bf 2} \
The self-consistent plots of $z$ vs. $t$ and $\tilde{e}$ vs. $t$ for fixed values of $g=0.031,~\lambda_1=0.004,~\lambda_2=0.01$ and different values of static JT coupling $g_1=0.145,~0.148,~0.152,~0.156$ and $0.159$.}

\vspace{0.2in}

\bigskip

{\vbox{
\begin{center}
      \epsfig{file=dyna7c.eps,width=8cm,height=6cm}
      \vspace{-1.0cm}
\end{center}
}}

\vspace{0.5cm}

{\small {\bf Fig.} \ {\bf 3} \
The self-consistent plots of $z$ vs. $t$ and $\tilde{e}$ vs. $t$ for fixed values of $g=0.031,~g_1=0.152,~\lambda_2=0.01$ and different values of normal EP coupling $\lambda_1=0.0,~0.004,~0.010,~0.015$ and $0.020$.}

The effect of the normal EP coupling on the SC order parameter and the
 static JT energy is shown in figure 3 for different values of the normal EP
 coupling $\lambda_1=0$ to $0.02$. On increase of the normal EP coupling, 
the JT energy gap decreases throughout the temperature range with the 
suppression of JT distortion temperature. However, the JT gap remains 
nearly constant at low temperatures for any change in the normal EP coupling. 
It is observed that the SC gap parameter is enhanced accompanied by the 
enhancement of the SC transition temperature with the increase of the normal
 EP coupling, but the SC gap nearly remains constant with the increase of 
EP coupling at low temperatures. Under these circumstances, the reduced gap
 parameter $2\Delta_0/k_BT_c$ decreases with increase of the normal EP 
coupling. Hence it is concluded that the normal EP coupling $\lambda_1$
 has profound effect on the SC transition temperature as well as the JT
 distortion temperature. It is further to note that the normal EP coupling
 can have $\lambda_1=f_1/2t_0$ from 0 to 0.02, which is small compared to the
 value of the EP coupling $\lambda_1=\frac{N(0)f_1^2}{\omega_0}$ lying 
between 0.10 to 0.15 for CDW superconductors\cite{mod2,res4,res5}. The 
effect of the normal EP coupling $\lambda_1$ in the transition temperatures is 
shown in figure 4. The JT distortion temperature decreases with increase of 
the normal EP coupling $\lambda_1$. On the other hand, the SC transition 
temperature increases monotonically with increase of the normal EP coupling 
$\lambda_1$. Hence it is concluded that the decrease of JT transition 
temperature enhances the SC transition temperature for the
 optimum value of $\lambda_1\simeq 0.0175$. When the normal electron-phonon
 coupling $(\lambda_1)$ increases, it strongly couples with at higher
 temperature $(t_{c}<t<t_{d})$, and it effectively reduces the static 
lattice energy $\tilde{e}$. In consequence there is the gain in electron
 energy. This results in enhancement of $t_{c}$. The existence of 
superconducting state at low temperature depends very sensitively on the 
strength of the EP coupling $(\lambda_{1})$ the effect of ionic 
size\cite{int16d} of iso-electronic M (Sr, Ca, Ba) ions on the 
interplay of superconductivity
 and structural transition can be understood as the consequence of higher
 EP coupling.
\vspace{0.2in}

\bigskip

{\vbox{
\begin{center}
      \epsfig{file=dyna13c.eps,width=8cm,height=6cm}
      \vspace{-1.0cm}
\end{center}
}}

\vspace{0.5cm}

{\small {\bf Fig.} \ {\bf 4} \
The plots of $t_c$ and $t_d$ vs. $\lambda_1$.}

The effect of the smaller values of the DJT coupling 
$\lambda_2(0-0.065)$ on the SC parameter and the JT energy gap is
 shown in figure 5. From figure 5 it is seen that the JT gap 
$\tilde{e}$ is suppressed throughout the temperature range with reduced
distortion temperature $t_d$ with the increase of the DJT coupling
 $\lambda_2$. The DJT coupling has no effect on the magnitude of the 
JT gap at low temperatures. On the other hand, with increase of the DJT
 coupling, the SC gap parameter is enhanced accompanied by the enhancement
 of the SC transition temperature. Again it is seen that the DJT\\
\vspace{0.2in}

\bigskip

{\vbox{
\begin{center}
      \epsfig{file=dyna8c.eps,width=8cm,height=6cm}
      \vspace{-1.0cm}
\end{center}
}}

\vspace{0.5cm}

{\small {\bf Fig.} \ {\bf 5} \
The self-consistent plots of $z$ vs. $t$ and $\tilde{e}$ vs. $t$ for fixed values of $g=0.031,~g_1=0.152,~\lambda_1=0.004$ and different values of DJT coupling $\lambda_2=0.0,~0.010,~0.025,~0.050$ and $0.065$.}\\
coupling 
has  no effect on the SC gap at the very low temperatures. 
On further increasing 
 the DJT coupling $\lambda_2=0.07$ to 0.20, the temperature 
dependent SC gap $z(t)$ and the JT gap $\tilde{e}(t)$ show different 
behavior as shown in figure 6. For DJT coupling $\lambda_2>0.07$, the JT 
distortion temperature becomes smaller than the SC transition temperature.
 Under this condition, the increase of the DJT coupling $\lambda_2$, the JT
 gap parameter is suppressed throughout the temperature range accompanied 
by the large suppression of the JT transition temperature, but the DJT 
coupling has no effect on the JT gap $\tilde{e}$ at low temperatures. It is 
noted here that the SC gap magnitude and the SC transition temperatures are 
unaffected for the ranges of higher values of $\lambda_2$. However, the SC 
gap magnitude are enhanced within the temperature ranges $0<t<t_c$. The 
reduced gap magnitude $2\Delta_0/k_BT_c\simeq 3.28$ remains constant for any 
change on the DJT coupling $\lambda_2$. The dependence of SC transition 
temperature $t_c$ and the distortion temperature $t_d$ on the DJT coupling 
$\lambda_2$ is shown clearly in figure 7. For lower values of DJT coupling
 $\lambda_2(0-0.063)$ the distortion temperature $t_d$ decreases rapidly 
with increase of $\lambda_2$, while the SC transition temperature increases
 slowly for all values of DJT coupling upto
 $\lambda_2=0.063$. On further increasing
 $\lambda_2$ to higher values $(\lambda_2>0.063)$, the SC transition
 temperature remains unaffected while the distortion temperature $t_d$ 
decreases continuously on increasing DJT coupling $\lambda_2$ as shown in
 figures 5 and 6. On further increasing $(\lambda_{2}>0.063)$, 
the modified static elastic 
energy becomes smaller than the electronic energy and the distortion
 temperature which is a measure of elastic energy, is suppressed as 
resulting in the situation $t_{s}< t_{c}$. Above the temperature $t_{d}$, 
all the electrons participate in the formation of Cooper Pairing and give 
rise to a constant $t_{c}$.
\vspace{0.2in}

\bigskip

{\vbox{
\begin{center}
      \epsfig{file=dyna9c.eps,width=8cm,height=6cm}
      \vspace{-1.0cm}
\end{center}
}}

\vspace{0.5cm}

{\small {\bf Fig.} \ {\bf 6} \
The self-consistent plots of $z$ vs. $t$ and $\tilde{e}$ vs. $t$ for fixed values of $g=0.031,~g_1=0.152,~\lambda_1=0.004$ and different values of DJT coupling $\lambda_2=0.07,~0.09,~0.12,~0.14$ and $0.20$.}
\vspace{0.2in}

\bigskip

{\vbox{
\begin{center}
      \epsfig{file=dyna14c.eps,width=8cm,height=6cm}
      \vspace{-1.0cm}
\end{center}
}}

\vspace{0.5cm}

{\small {\bf Fig.} \ {\bf 7} \
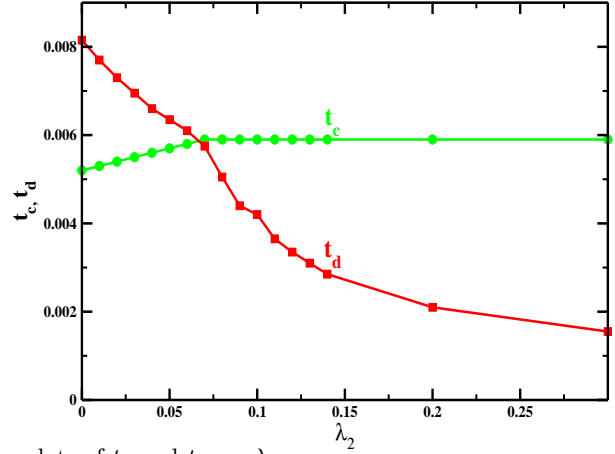
The plots of $t_c$ and $t_d$ vs. $\lambda_2$.}

The effect of the reduced external frequency $c$ on the SC gap parameter 
and the JT energy gap is shown in figure 8. With increase in frequency from
 $c=0.10$ to 0.96, the magnitude of the JT gap is suppressed considerably 
accompanied by the suppression of the distortion temperature $t_d$, but the 
gap magnitude at low temperatures remains unaffected with increase in 
frequency. On the other hand with increase of the frequency, the SC gap 
parameter is enhanced along with the SC transition temperature; but the 
magnitude of the SC gap remains unaffected at low temperatures. The change 
of the distortion temperature $t_d$ and SC transition temperature $t_c$ with 
the change of the external frequency is shown in figure 9. These two
 transition temperatures show different types of 
behavior for the lower external\\
\bigskip

{\vbox{
\begin{center}
      \epsfig{file=dyna6c.eps,width=8cm,height=6cm}
      \vspace{-1.0cm}
\end{center}
}}

\vspace{0.5cm}

{\small {\bf Fig.} \ {\bf 8} \
The self-consistent plots of $z$ vs. $t$ and $\tilde{e}$ vs. $t$ for fixed values of $g=0.031,~g_1=0.152,~\lambda_1=0.004,~\lambda_2=0.01$ and different values of the external frequencies $c=0.10,~0.70,~0.80,~0.90$ and $0.96$.}\\
\bigskip

{\vbox{
\begin{center}
      \epsfig{file=dyna12c.eps,width=8cm,height=6cm}
      \vspace{-1.0cm}
\end{center}
}}

\vspace{0.5cm}

{\small {\bf Fig.} \ {\bf 9} \
The plots of $t_c$ and $t_d$ vs. $c$.}\\
frequencies and higher external frequencies upto the frequency $c\simeq 1$.
 For lower frequencies, the distortion temperature $t_d$ decreases;
 while the SC transition temperature 
slowly increases. On further increasing the frequency, the distortion 
temperature $t_d$ suddenly increases, then decreases rapidly and then
 remains constant for higher frequencies. On the other hand, on increasing 
the frequency to higher values, the SC transition temperature suddenly drops 
to a lower value, then slowly increases and then remains constant for higher
 frequencies.  Figures 8 and 9 exhibit the influence of sound 
frequency ($c$) on the interplay of the SC and JT strain parameters. 
With the increase of the sound frequency, the coupling of the phonons to 
the band electrons becomes stronger i.e., the normal EP coupling 
$(\lambda_{1})$ and the DJT coupling $(\lambda_{2})$ become stronger. 
As discussed for $\lambda_{1}$ and $\lambda_{2}$, we will find that the 
increase of phonon frequency will reduce the static lattice energy resulting 
 in the suppression of distortion temperature $t_{s}$ and 
enhancement of $t_{c}$.\\

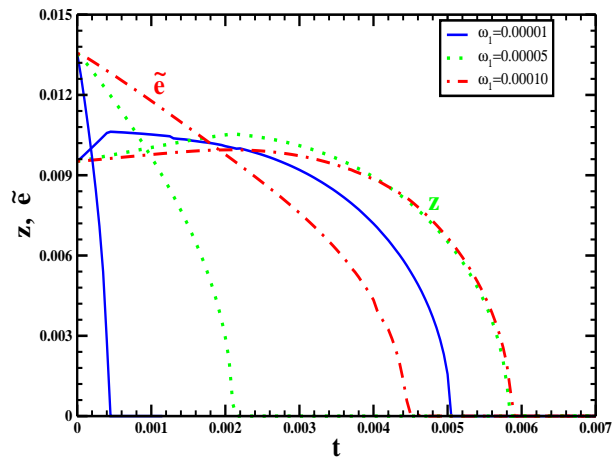
\bigskip

{\vbox{
\begin{center}
      \epsfig{file=dyna16c.eps,width=8cm,height=6cm}
      \vspace{-1.0cm}
\end{center}
}}

\vspace{0.5cm}

{\small {\bf Fig.} \ {\bf 10} \
The self-consistent plots of $z$ vs. $t$ and $\tilde{e}$ vs. $t$ for fixed values of $g=0.031,~g_1=0.152,~\lambda_1=0.004,~\lambda_2=0.01$ and  different lower phonon vibrational frequencies $\omega_1=0.00001,~0.00005,~0.00010$ and $0.00020$.}\\

\bigskip

{\vbox{
\begin{center}
      \epsfig{file=dyna5c.eps,width=8cm,height=6cm}
      \vspace{-1.0cm}
\end{center}
}}

\vspace{0.5cm}

{\small {\bf Fig.} \ {\bf 11} \
The self-consistent plots of $z$ vs. $t$ and $\tilde{e}$ vs. $t$ for fixed values of $g=0.031,~g_1=0.152,~\lambda_1=0.004,~\lambda_2=0.01$ and different higher vibration frequencies $\omega_1=0.0002,~0.0004,~0.0008$ and $0.0100$.}

Figure 10 shows the temperature dependence of the SC gap  
and the JT gap parameters for different values of the phonon
 vibrational frequency. With
 the increase of phonon vibrational frequency $\omega_1$, 
 the JT gap increases throughout
 the temperature range and simultaneously the JT distortion temperature is 
also enhanced. This shows that the phonon vibrational frequency enlarges the JT 
gap. It is to note further that, in the interplay region of the lattice strain
 and the SC order, the SC parameter is suppressed at low temperatures. However,
 for the increase of the low vibrational frequencies the SC gap parameter is 
enhanced accompanied by the enhancement of its transition temperature. The 
effect of higher phonon vibrational frequency on the gap parameters is shown in
 figure 11. With the increase of the phonon vibrational 
 frequency $\omega_1$, the JT gap 
is enhanced throughout the temperature range along with its transition 
temperature; but the JT gap magnitude, at very low temperature remains 
unaffected with $\omega_1$. On the other hand, the SC gap parameter is reduced
 throughout the temperature range with increase of the phonon frequency. 
Though the SC transition temperature is reduced with the phonon frequency, the 
magnitude of the SC gap remains unaffected at lower temperatures.\\
\bigskip

{\vbox{
\begin{center}
      \epsfig{file=dyna11c.eps,width=8cm,height=6cm}
      \vspace{-1.0cm}
\end{center}
}}

\vspace{0.5cm}

{\small {\bf Fig.} \ {\bf 12} \
The plots of $t_c$ and $t_d$ vs. $\omega_1$.}\\
 
The 
dependence of the SC transition temperature and the JT distortion temperature on
 the phonon vibrational frequency $\omega_1$ is shown in figure 12. For lower
 frequencies upto $\omega_1=0.0013$, both the SC 
 transition temperature $t_c$ as 
well as the JT distortion temperature $t_d$
  increase with increase of the phonon 
vibrational frequency. On further increasing phonon vibrational frequency 
$\omega_1$, the distortion temperature increases and remains constant for
 higher vibrational frequencies. Correspondingly the SC transition temperature 
decreases with the higher vibrational frequencies and then remains constant.
 Thus the $t_c$ and $t_d$ of the system are not affected for higher vibrational
 frequencies. The interplay between superconductivity and lattice
 distortion has been distinctively observed in 
$La_{2-x} M_{x} CuO_{4}$ $[M=Sr, Ca, Ba]$\cite{int16d}. 
The Ca-substituted compound has higher value of the electron-lattice 
coupling constant than that of Sr-substituted system. The smaller is the
 ionic size, the larger will be the phonon vibration in the system. 
For lower phonon frequencies, both the lattice distortion temperature
 $t_{s}$ and SC transition temperature $t_{c}$ are enhanced\cite{int16d}.
 For high frequency 
phonons the lattice distortion is enhanced resulting in the 
suppression $t_{c}$.
\section{Conclusion}
  We report here a model study to treat the static and dynamic 
Jahn-Teller effects on high temperature copper oxide superconductors. 
The static JT effects are considered as the renormalization to the energy
 of the doubly degenerate conduction electron bands. Further, the dynamic
 JT effects renormalize both the conduction bands as well as the 
superconducting gap parameter under lattice strain through the
 electron-phonon coupling constant.
               The solutions are obtained by using the Zubarev double time
 Green's function methods including the higher order mixing Green's
 functions within mean-field approximations. We observe the changes in
 lattice strain, when the temperature decreases to below SC transition
 temperature $t_{c}$, as observed by the calculation of 
Ghosh $et~el.$\cite{int29}.
 Further, we observe here some new results beyond mean-field approximation 
in a generalized model.

In the model calculation for the gap equation for the 
high-T$_c$ cuprates, we have considered the Jahn-Teller type lattice strain as 
a pseudogap interacting strongly with the superconducting pairing 
amplitudes. The static lattice strain suppresses the 
superconducting gap as reported by the neutron scattering studies
and the theoretical calculations\cite{int29,int30,mod1,mod2}.
 Keeping in mind the important role played by the PG, we consider 
here the DJT effect on the superconducting 
gap in high-T$_c$ systems.
In addition to the static band Jahn-Teller effect, we incorporate 
the normal phonon interaction to the electron densities of the degenerate 
conduction band and the DJT interaction to the difference in 
electron densities of the JT splitted two bands. The superconducting gap 
equation is calculated by the Zubarev's technique of Green's functions and the 
modified BCS gap equation is calculated from the modified conduction band 
energy $\tilde{\epsilon}_{\alpha k}$ and the modified SC gap
 parameter $\tilde{\Delta}_\alpha$. The lattice strain is calculated by
 minimizing the free energy. The SC gap and JT energy gap equations are 
solved self-consistently. It is observed that the SC gap and the SC transition 
temperature are enhanced by the DJT interaction as compared to the 
static case. The normal EP coupling suppresses the JT distortion 
temperature and enhances the SC transition temperature. Similarly,
 for smaller DJT coupling $(\lambda_2<0.075)$, the JT distortion
 temperature is suppressed, while the SC transition temperature is enhanced.
 For higher values of the DJT coupling $(\lambda_2>0.075)$, the 
JT distortion is suppressed, but the SC transition is not enhanced further.
 It is found that the external frequency suppresses the JT transition 
temperature but enhances the SC transition temperature for frequencies
 $c\leq 1$. It is further noticed that both the JT distortion temperature 
and the SC transition temperature are enhanced for the lower values of the
phonon vibrational frequencies, while both the transition temperatures are not
 affected by the higher phonon vibrational frequencies. The electronic 
correlation effects and the $d$-wave symmetry are not discussed 
 in the present model calculations. The present model 
study helps in understanding the interesting phenomenon of the
 interplay of dynamic structural transition with the superconductivity
 in cuprate systems. However, it will be interesting to study, how the
 antiferromagnetism spin fluctuations leading to $d$-wave pairing 
superconductivity will be influenced by the dynamic structural distortion. 
This will be the subject matter for a future study.

\section*{Acknowledgements}
The authors gracefully acknowledge the research facilities offered
 by the Institute of Physics, Bhubaneswar, India during their short stay.

\end{document}